\documentclass[epjCONF]{svjour}
\usepackage{graphicx}
\usepackage[varg]{txfonts} 
\usepackage[latin1]{inputenc}
\session-title{%
19$^{\textnormal{\footnotesize th}}$ International %
IUPAP Conference on Few-Body Problems in Physics%
}
\begin{document}
\title{%
Photodisintegration of $^3H$ in a three dimensional faddeev approach
}%
\author{%
S. Bayegan\inst{1}\fnmsep\thanks{\email{bayegan@khayam.ut.ac.ir}}
\and %
M. A. Shalchi\inst{1}
\and %
M. R. Hadizadeh\inst{2} }
\institute{%
 Department of Physics, University of Tehran, P.O.Box
14395-547, Tehran, Iran
\and %
Instituto de F\'{i}sica Te\'{o}rica, Universidade Estadual Paulista,
Rua Dr. Bento Teobaldo Ferraz 271, Bl. II, Barra Funda, 01140-070,
S\~{a}o Paulo, Brazil
}
\abstract{ An interaction of a photon with $^3H$ is invstigated
based on a three dimensional Faddeev approach. In this approach the
three-nucleon Faddeev equations with two-nucleon interactions are
formulated with consideration of the magnitude of the vector Jacobi
momenta and the angle between them with the inclusion of the
spin-isospin quantum numbers, without employing a partial wave
decomposition. In this formulation the two body t-matrices and
triton wave function are calculated in the three dimensional
approach using AV18 potential. In the first step we use the standard
single nucleon  current in this article.
} 
\maketitle
%
%
%
\section{Introduction}
\label{SchmidtPL_intro}

Since the early days of the study of the nuclear physics so many
efforts have been performed on 3N systems considering real or
virtual photon interactions
\cite{Wigner_PhR43}-\cite{Gerjuoy_PhR61}. Also several studies on
the behavior of 3N bound states in real or virtual photon absorbtion
have been reported\cite{Collard_PhR11}-\cite{Schiff_Phl11}. Before
sixties variational approach was used  for these calculations and
works using this approach are still continuing. After introducing
Faddeev formulation for three body systems
\cite{Faddeev_ThF39}-\cite{Alt_NPB2}, new efforts using this scheme
were started. As an example one can point out the early calculations
of electrodisintegration \cite{Lehman_PhRL23} and
photodisintegration\cite{Barbour_PhRL19} with $^3He$ and $^3H$. An
improvement in  the photodisintegration calculation of the bound and
3N continuum  with the same 3N hamiltonian have been performed
\cite{Gibson_PhRC11}. There are also other approaches to calculate
electromagnetic interactions with light nuclei such as
Green-function-Monte-Carlo method \cite{Carlson_PhRC36},
hyperspherical harmonic expansions \cite{Viviani_PhRC61}, and
Lorentz integral transform (LIT) method \cite{Efros_PhLB338}.There
is a very good review of Faddeev calculations on the interaction of
real or virtual photon with $^3He$  \cite{golak}. In this work like
previous calculations the partial wave decomposition has been used.
In PW approach one should sum all PW to maximum angular momentum
where the calculation is converged. The problem is that in higher
energies this maximum angular momentum increases and we should solve
more complicated equations. To avoid this complexity one should use
vector momentum as basis states \cite{Rice_FB14}. To this aim in the
past decade the main steps have been taken by Ohio-Bochum
collaboration (Elster, Gl\"{o}ckle et al.) and Bayegan et al. to
implement the 3D approach in few-body bound and scattering
calculations (see for examples Refs.
\cite{Fachruddin_PhRC62}-\cite{Bayegan_NPA814}). It should be clear
that the building blocks to the few-body calculations without
angular momentum decomposition are two-body off-shell t-matrices,
which depend on the magnitudes of the initial and final Jacobi
momenta and the angle between them. Fachruddin et al. have
calculated the NN bound and scattering states in a 3D representation
using  the Bonn-B and the AV18 potentials
\cite{Fachruddin_PhRC62}-\cite{Fachruddin_PhRC63}. Recently there
has been efforts to do the same calculation using chiral potential
\cite{Bayegan_PhRC79}. our aim in this work is to formulate
photodisintegration of $^3H$ in a three dimensional Faddeev
approach. In the first step we ignore three body forces and we just
use the single nucleon current. We will use AV18 potential and
triton wave function which has been calculated in our previous work
\cite{Bayegan_PhRC77}.

This manuscript has been organized as follow: in section
\ref{integral equation} we explain our basic states and we evaluate
all of the matrix elements in these basis. In section
\ref{singularity} we introduce our singularity problem and its
solution. We finish in section \ref{summary} with a summary and
outlook.

\section{Integral equation of nuclear matrix elements without partial wave decomposition  }
\label{integral equation}

 To calculate the photodisintegration
observable we first need to calculate nuclear matrix elements in the
Faddeev scheme. For more details see Ref.\cite{golak}.

\begin{equation}
  \label{SchmidtPL_eq:2}
N=\frac{1}{2}\langle\phi_0|(1+tG_0)P|U\rangle
\end{equation}

\begin{equation}
  \label{SchmidtPL_eq:1}
|U\rangle=(1+P)J|\psi\rangle+tG_0P|U\rangle
\end{equation}

In above equations $t$ is NN t-operator which obeys
Lipmann-Schwinger equations, $G_0$ is free propagator, $P$ is
permutation operator, $|\psi\rangle$ is three body bound state and
$|U\rangle$ is an axillary state. Three body forces have been
ignored. $|\phi_0\rangle$ is a subsection of the fully antisymmetric
free state, $|\Phi_0\rangle$, in which nucleons 2 and 3 are in
subsystem.

\begin{equation}
  \label{SchmidtPL_eq:3}
|\Phi_0\rangle=(1+P)|\phi_0\rangle
\end{equation}

$|\phi_0\rangle$ is also our basic state to solve the integral
equation (2) and is antisymmetric under permutation of nucleons 2
and 3.

\begin{equation}
  \label{SchmidtPL_eq:5}
|\phi_0 \rangle \equiv|\textbf{p}\textbf{q}m_1m_2m_3\nu_1\nu_2\nu_3
\rangle^a
\end{equation}

In equation (4) $\textbf{p}$ and $\textbf{q}$ are jacobi momenta and
m's and $\nu$'s are the spin and isospin of the individual nucleons
respectively.

 Orthonormality and completeness relations of these
basic states can be considered as bellow:

\begin{eqnarray}
  \label{SchmidtPL_eq:1}
&&^a\langle\textbf{p}\textbf{q}m_1m_2m_3\nu_1\nu_2\nu_3|\textbf{p}'\textbf{q}'m'_1m'_2m'_3\nu'_1\nu'_2\nu'_3
\rangle^a \nonumber
\\&&=\frac{1}{2}\{\delta(\textbf{p}-\textbf{p}')\delta_{m_2m'_2}\delta_{m_3m'_3}\delta_{\nu_2\nu'_2}\delta_{\nu_3\nu'_3}\nonumber\\
&&-
\delta(\textbf{p}+\textbf{p}')\delta_{m_2m'_3}\delta_{m_3m'_2}\delta_{\nu_2\nu'_3}\delta_{\nu_3\nu'_2}\}\delta(\textbf{q}-\textbf{q}')\delta_{m_1m'_1}\delta_{\nu_1\nu'_1}
\end{eqnarray}

\begin{eqnarray}
  \label{SchmidtPL_eq:1}
\sum_{\begin{array}{ccc}
                  m_1 & m_2 & m_3 \\
                  \nu_1 &\nu_2 & \nu_3
                \end{array}}\int
                d^3\textbf{p}\,d^3\,\textbf{q}|\textbf{p}\textbf{q}m_1m_2m_3\nu_1\nu_2\nu_3 \rangle^a
                \nonumber \\\times ^a\langle\textbf{p}\textbf{q}m_1m_2m_3\nu_1\nu_2\nu_3|\equiv1
\end{eqnarray}
Considering these properties we can rewrite the integral equations
(1) and (2) in our basic stats.
\begin{eqnarray}
  \label{SchmidtPL_eq:1}
N&&=\frac{1}{2}\,^a\langle\textbf{p}\textbf{q}m_1m_2m_3\nu_1\nu_2\nu_3|(1+tG_0)P|U
\rangle\nonumber
\\&&=\frac{1}{2}\,^a\langle\textbf{p}\textbf{q}m_1m_2m_3\nu_1\nu_2\nu_3|P|U \rangle\nonumber
\\&&+\frac{1}{2}\,^a\langle\textbf{p}\textbf{q}m_1m_2m_3\nu_1\nu_2\nu_3|tG_0P|U \rangle
\end{eqnarray}
\begin{eqnarray}
  \label{SchmidtPL_eq:1}
&&^a\langle\textbf{p}\textbf{q}m_1m_2m_3\nu_1\nu_2\nu_3|U \rangle\nonumber \\
&&=^a\langle\textbf{p}\textbf{q}m_1m_2m_3\nu_1\nu_2\nu_3|(1+P)J|\psi
\rangle \nonumber
\\&&+^a\langle\textbf{p}\textbf{q}m_1m_2m_3\nu_1\nu_2\nu_3|tG_0P|U \rangle
\end{eqnarray}

The effect of permutation operator on our basic states can be
considered as follow:
\begin{eqnarray}
  \label{SchmidtPL_eq:1}
&&\langle\textbf{p}\textbf{q}m_1m_2m_3\nu_1\nu_2\nu_3|P|\textbf{p}'\textbf{q}'m'_1m'_2m'_3\nu'_1\nu'_2\nu'_3
\rangle\nonumber
\\&&= \delta(\textbf{p}+\frac{1}{2}\textbf{p}'+\frac{3}{4}\textbf{q}')
\delta(\textbf{q}-\textbf{p}'+\frac{1}{2}\textbf{q}')\nonumber
\\ &&\times\delta_{m_1m'_2}\delta_{m_2m'_3}\delta_{m_3m'_1}
\delta_{\nu_1\nu'_2}\delta_{\nu_2\nu'_3}\delta_{\nu_3\nu'_1}\nonumber
\\ &&+\delta(\textbf{p}+\frac{1}{2}\textbf{p}'-\frac{3}{4}\textbf{q}')
\delta(\textbf{q}+\textbf{p}'+\frac{1}{2}\textbf{q}')\nonumber
\\ &&\times\delta_{m_1m'_3}\delta_{m_2m'_1}\delta_{m_3m'_2}
\delta_{\nu_1\nu'_3}\delta_{\nu_2\nu'_1}\delta_{\nu_3\nu'_2}\nonumber
\\ &&=\delta(\textbf{p}+\pi_2)\delta(\textbf{p}'-\pi_1)\nonumber
\\ &&\times\delta_{m_1m'_2}\delta_{m_2m'_3}\delta_{m_3m'_1}
\delta_{\nu_1\nu'_2}\delta_{\nu_2\nu'_3}\delta_{\nu_3\nu'_1}\nonumber
\\ &&+\delta(\textbf{p}-\pi_2)\delta(\textbf{p}'+\pi_1)\nonumber
\\ &&\times\delta_{m_1m'_3}\delta_{m_2m'_1}\delta_{m_3m'_2}
\delta_{\nu_1\nu'_3}\delta_{\nu_2\nu'_1}\delta_{\nu_3\nu'_2}
\end{eqnarray}

Where:
\begin{eqnarray}
  \label{SchmidtPL_eq:1}
\pi_1=\textbf{q}+\frac{1}{2}\textbf{q}'~~~~~~\pi_2=\frac{1}{2}\textbf{q}+\textbf{q}'
\end{eqnarray}
Now with respect to above relation and symmetry considerations  we
can evaluate equations (7) and(8) as follow:
\begin{eqnarray}
  \label{SchmidtPL_eq:1}
&&N=\frac{1}{2}\{\langle-\frac{1}{2}\textbf{p}-\frac{3}{4}\textbf{q},\,\textbf{p}-\frac{1}{2}\textbf{q}\,m_2m_3m_1\nu_2\nu_3\nu_1|U
\rangle\nonumber
\\&& \langle-\frac{1}{2}\textbf{p}+\frac{3}{4}\textbf{q},\,-\textbf{p}-\frac{1}{2}\textbf{q}\,m_3m_1m_2\nu_3\nu_1\nu_2|U \rangle\}\nonumber
\\&&+\sum_{\begin{array}{cc}
          m'_2 & m'_3 \\
          \nu'_2 & \nu'_3
        \end{array}}\int d^3\textbf{q}
        ^a\langle\textbf{p}\,m_2m_3\nu_2\nu_3|t|\frac{1}{2}\textbf{q}+\textbf{q}',m'_2m'_3\nu'_2\nu'_3 \rangle^a\nonumber
        \\&&\frac{1}{E-\frac{\textbf{q}^2+\textbf{q}'^2+\textbf{q}\cdot\textbf{q}'}{m}}\langle-\frac{1}{2}\textbf{q}'-\textbf{q},\,\textbf{q}'\,m'_2m'_3m_1\nu'_2\nu'_3\nu_1|U \rangle
\end{eqnarray}

\begin{eqnarray}
  \label{SchmidtPL_eq:1}
&&\langle\textbf{p}\textbf{q},\,m_1m_2m_3\nu_1\nu_2\nu_3|U
\rangle\nonumber
\\&&=\langle\textbf{p}\textbf{q},\,m_1m_2m_3\nu_1\nu_2\nu_3|(1+P)J|\psi \rangle\nonumber
\\&&+\sum_{\begin{array}{cc}
          m'_2 & m'_3 \\
          \nu'_2 & \nu'_3
        \end{array}}\int d^3\textbf{q}
        ^a\langle\textbf{p}\,m_2m_3\nu_2\nu_3|t|\frac{1}{2}\textbf{q}+\textbf{q}',m'_2m'_3\nu'_2\nu'_3 \rangle^a\nonumber
        \\&&\frac{1}{E-\frac{\textbf{q}^2+\textbf{q}'^2+\textbf{q}\cdot\textbf{q}'}{m}}\langle-\frac{1}{2}\textbf{q}'-\textbf{q},\,\textbf{q}'\,m'_2m'_3m_1\nu'_2\nu'_3\nu_1|U \rangle
\end{eqnarray}

The firs term in the equation(12) can be evaluated as:
\begin{eqnarray}
  \label{SchmidtPL_eq:1}
&&^a\langle\textbf{p}\textbf{q},\,m_1m_2m_3\nu_1\nu_2\nu_3|(1+P)J|\psi
\rangle\nonumber
\\&&=\sum_{m',\nu'}\int d^3\textbf{p}'d^3\textbf{q}'\nonumber \\&&^a\langle\textbf{p}\textbf{q},\,m_1m_2m_3\nu_1\nu_2\nu_3|(1+P)J|\textbf{p}'\textbf{q}',\,m'_1m'_2m'_3\nu'_1\nu'_2\nu'_3 \rangle^a\nonumber
\\&& \times ^a\langle\textbf{p}'\textbf{q}',\,m'_1m'_2m'_3\nu'_1\nu'_2\nu'_3|\psi \rangle
\end{eqnarray}
Now we concentrate on the elements of these equations i.e. current,
two-body t-matrix and triton wave function, more precisely.

\subsection{current}
\label{current:21}

Considering the symmetry properties we have:
\begin{eqnarray}
  \label{SchmidtPL_eq:1}
&&^a\langle\textbf{p}\textbf{q},\,m_1m_2m_3\nu_1\nu_2\nu_3|(1+P)J^{SN}|\psi
\rangle\nonumber
\\&&=3^a\langle\textbf{p}\textbf{q},\,m_1m_2m_3\nu_1\nu_2\nu_3|(1+P)J^{SN}(1)|\psi\rangle\nonumber\\
\end{eqnarray}
Matrix elements of single nucleon current can be evaluated as
follow:
\begin{eqnarray}
  \label{SchmidtPL_eq:1}
&&^a\langle\textbf{p}\textbf{q},\,m_1m_2m_3\nu_1\nu_2\nu_3|J(1)|\textbf{p}'\textbf{q}',\,m'_1m'_2m'_3\nu'_1\nu'_2\nu'_3
\rangle\nonumber
\\&&=\delta(\textbf{q}'-\textbf{q}+\frac{2}{3}\textbf{Q})\nonumber \\ &&\times\frac{1}{2}
[\delta(\textbf{p}-\textbf{p}')\delta_{m_2m'_2}\delta_{m_3m'_3}\delta_{\nu_2\nu'_2}\delta_{\nu_3\nu'_3}\nonumber
\\
&&~~~~~~-\delta(\textbf{p}+\textbf{p}')\delta_{m_2m'_3}\delta_{m_3m'_2}\delta_{\nu_2\nu'_3}\delta_{\nu_3\nu'_2}]\times
J_{\begin{array}{cc}
    m_1 & m'_1 \\
    \nu_1 & \nu'_1
  \end{array}}(\textbf{Q},\textbf{q})\nonumber\\
\end{eqnarray}
In above equation $\textbf{Q}$ is the momentum of photon. We need to
rewrite the single nucleon current operator in a form which is
suitable for our basic states. The current operator which we will
use is:
\begin{eqnarray}
  \label{SchmidtPL_eq:1}
J=G_E(Q)\frac{\textbf{k}_1+\textbf{k}'_1}{2m_N}+\frac{i}{2m_N}G_M(Q)\sigma
\times (\textbf{k}_1-\textbf{k}'_1)
\end{eqnarray}
Which is summation of convection current and spin current. $G_E(Q)$
and $G_M(Q)$ are electric and magnetic form factors respectively.
For the convection part we have:
\begin{eqnarray}
\label{SchmidtPL_eq:1} \textbf{k}_1+\textbf{k}'_1=2
\textbf{q}+\textbf{Q}+\frac{2}{3}\textbf{K}
\end{eqnarray}
As we will show we have to choose coordinate system in which the $z$
axis is along the $Q$ vector and we also need tensor component of
current so the second and the third terms of the right hand side of
equation(17) will vanish. Thus for the convection current we have:
\begin{eqnarray}
  \label{SchmidtPL_eq:1}
J_{\pm1}^{convec}=G_E(Q)\frac{q_{\pm1}}{m_N}
\end{eqnarray}

And the tensor component of spin part can also be evaluated as:

\begin{eqnarray}
  \label{SchmidtPL_eq:1}
J_{\pm1}^{Spin}=\frac{-\sqrt{2}Q}{2m_N}G_M(Q)S_{\pm}
\end{eqnarray}

\subsection{two-body t- matrix}
\label{two body:22} Two body t-matrices can be related to the one
which calculated in helicity basis:
\begin{eqnarray}
  \label{SchmidtPL_eq:1}
&&^a\langle\textbf{p}m_1m_2n\nu_1\nu_2|t|\textbf{p}'m_1m_2\nu'_1\nu'_2\rangle^a=
\frac{1}{4}\delta_{(\nu_1+\nu_2),(\nu'_1+\nu'_2)}\nonumber \\
&&e^{i(\Lambda_0\phi_p-\Lambda'_0\phi_p')}\sum_{\pi
st}(1-\eta_{\pi})C(\frac{1}{2}\frac{1}{2}t,\nu_1\nu_2)
C(\frac{1}{2}\frac{1}{2}t,\nu'_1\nu'_2)\nonumber
\\
&&C(\frac{1}{2}\frac{1}{2}S,m_1m_2\Lambda_0)C(\frac{1}{2}\frac{1}{2}S,m'_1m'_2
\Lambda'_0)\nonumber\\&&\sum_{\Lambda\Lambda'}d_{\Lambda_0\Lambda}^S(\theta_p)
d_{\Lambda'_0\Lambda'}^S(\theta_p')\frac{\sum
e^{iN(\phi_p-\phi_p')
}d_{N\Lambda}^S(\theta_p)d_{N\Lambda'}^S(\theta_p')}{d_{\Lambda'\Lambda}^S(\theta_{pp'})}\nonumber
\\&&t_{\Lambda\Lambda'}^{\pi S t}(p,p',\cos\theta_{pp'},z)
\end{eqnarray}
In the above relation $z=E-\frac{3q^2}{4m}$ is the energy of
subsystem. As we know two-body function has a singularity in the
energy of deuteron, $z=E_d$. To remove this singularity we should
consider t-operator as follow:

\begin{eqnarray}
\label{SchmidtPL eq:1}
&&^a\langle\textbf{p}m_1m_2n\nu_1\nu_2|t|\textbf{p}'m_1m_2\nu'_1\nu'_2\rangle^a\nonumber\\
&&=\frac{^a\langle\textbf{p}m_1m_2n\nu_1\nu_2|\hat{t}|\textbf{p}'m_1m_2\nu'_1\nu'_2\rangle^a}{z-E_d}
\end{eqnarray}

Two body t-matrix in helicity basis has been calculated
before\cite{Fachruddin_PhRC62}.

\begin{eqnarray}
  \label{SchmidtPL_eq:1}
 &&t_{\Lambda '\Lambda }^{\pi St}\left( {p',p,\cos\theta } \right) = V_{\Lambda '\Lambda }^{\pi St}\left( {p',p,\theta } \right)\nonumber \\
 {\rm{                   }} +&& \frac{1}{2}\int {dp''{{p''}^2}} \int\limits_{ - 1}^1 {d\left( {\cos \theta ''} \right)v_{\Lambda '1}^{\pi St,\Lambda }\left( {p',p'',\theta ',\theta ''} \right)} {G_0}\left( {p''}\right)\nonumber \\ &&t_{1\Lambda }^{\pi St}\left( {p'',p,\theta ''} \right) \nonumber \\
 {\rm{                   }} + &&\frac{1}{2}\int {dp''{{p''}^2}} \int\limits_{ - 1}^1 {d\left( {\cos \theta ''} \right)v_{\Lambda '0}^{\pi St,\Lambda }\left( {p',p'',\theta ',\theta ''} \right)} {G_0}\left( {p''} \right)\nonumber \\&&t_{0\Lambda }^{\pi St}\left( {p'',p,\theta ''} \right)
\end{eqnarray}

Where
\begin{eqnarray}
  \label{SchmidtPL_eq:1}
 v_{\Lambda '\Lambda ''}^{\pi St,\Lambda } (p',p'',\theta ',\theta '') = \int\limits_0^{2\pi } {d\phi ''{e^{ - i\Lambda \left( {\phi ' - \phi ''} \right)}}V_{\Lambda '\Lambda ''}^{\pi St}\left( {\vec p',\vec p''}
 \right)}\nonumber \\
 \end{eqnarray}

\subsection{triton wave function}
\label{triton:23}

For evaluating the Triton wave function we need to make a relation
between this wave function in our basic states to the one which has
been calculated in the following basis\cite{Bayegan_PhRC77}:
\begin{eqnarray}
  \label{SchmidtPL_eq:1}
\langle\textbf{p}\textbf{q}(s\frac{1}{2})Sm_S(t\frac{1}{2})Tm_T|\psi\rangle
=\langle pq, X_{pq}, \alpha|\psi\rangle
\end{eqnarray}
We can relate these states to our free spin and isospin states with
Clebsch-Gordan coefficients.
\begin{eqnarray}
  \label{SchmidtPL_eq:1}
g_{\gamma\alpha}=\langle\gamma|\alpha\rangle
\end{eqnarray}
Where
\begin{eqnarray}
  \label{SchmidtPL_eq:1}
&&|\alpha\rangle=|(s\frac{1}{2})Sm_S,(t\frac{1}{2})Tm_T\rangle
\nonumber
\\&&|\gamma\rangle=|m_1m_2m_3\nu_1\nu_2\nu_3\rangle
\end{eqnarray}
It is very important to mention that the spin of the nucleons is
quantized in the direction of the $z$ axis which in the calculation
of wave function it has been chosen to be in the direction of
$\textbf{q}$. But we have to consider the $z$ axis along the
direction of incident photon $\textbf{Q}$. So we should first rotate
the spin of the nucleons in our basis to be settled in the direction
of $\textbf{q}$ axis. Then we should use Clebsch-Gordan coefficients
to obtain the wave function in the calculated basis mentioned in the
equation (24):
\begin{eqnarray}
  \label{SchmidtPL_eq:1}
&&\langle\textbf{p}\textbf{q}m_1m_2m_3\nu_1\nu_2\nu_3|\psi\rangle
 \nonumber
\\ &&=\sum_{m'_1m'_2m'_3}\sum_{\alpha} D_{m_1m'_1}(\theta_q,\phi_q) D_{m_2m'_2}
(\theta_q,\phi_q)
D_{m_3m'_3}(\theta_q,\phi_q)g_{\gamma\alpha}\nonumber\\ &&\times
\langle pq, X_{pq}, \alpha|\psi\rangle
\end{eqnarray}

\section{Singularity problem}
\label{singularity}

In order to consider the singularity problem we can rewritten the
equation (11) an (12) in a unified form ignoring isospin
 dependent which is similar to spin dependent.
\begin{eqnarray}
\label{Schmidth eq:1}
&&U_{m_1m_2m_3}(\textbf{p},\textbf{q},\textbf{Q})=U'_{m_1m_2m_3}(\textbf{p},\textbf{q},\textbf{Q})\nonumber \\
&&+\sum_{m'_2m'_3}\int
d^3q''\frac{U_{m'_2m'_3m_1}(\pi_2,\textbf{q}'',\textbf{Q})}{E-\frac{q^2+q''^2-\textbf{q}\cdot\textbf{q}''}{m}}\frac{\hat{t}^a_{m_2m_3m'_3m'_3}(\textbf{p},\pi_1,z)}{E+i\epsilon-E_d-\frac{3q^2}{4m}}\nonumber
\\
\end{eqnarray}
To solve this integral equation we should evaluate singularity in
the denominator of the propagator which is a function of $q''$ and
angle between $q''$and $q$. So instead of singular point we have a
region of singularity in $q-q''$ plane. There is a solution to this
moving singularity in Ref.\cite{Liu_PhRC72}. For using this method
we have to put $z$ axis along the $\textbf{q}$. But because of
simplification in current operator and final cross section we should
choose the $z$ axis in the direction of the momentum of the photon,
$\textbf{Q}$. So in order to evaluate the singularity we should use
another method which is introduced in Ref.\cite{Elster_FbS45}.
Therefore one should separate angle part of delta functions as
follow:

\begin{eqnarray}
\label{Schmidth eq:1}
\delta(\textbf{p}'+\pi_1)\delta(\textbf{p}''-\pi_1)=\frac{\delta(p'-\pi_1)}{p'^2}\frac{\delta(p''-\pi_1)}{p''^2}\nonumber
\\\delta(\hat{\textbf{p}}'+\hat{\pi_1})\delta(\hat{\textbf{p}}''-\hat{\pi_1})\nonumber
\\
\delta(\textbf{p}'-\pi_1)\delta(\textbf{p}''+\pi_1)=\frac{\delta(p'-\pi_1)}{p'^2}\frac{\delta(p''-\pi_1)}{p''^2}\nonumber
\\\delta(\hat{\textbf{p}}'-\hat{\pi_1})\delta(\hat{\textbf{p}}''+\hat{\pi_1})
\end{eqnarray}
And then the integral equation can be rewrite as follow:

\begin{eqnarray}
\label{Schmidth eq:1}
&&U_{m_1m_2m_3}(\textbf{p},\textbf{q},\textbf{Q})=U'_{m_1m_2m_3}(\textbf{p},\textbf{q},\textbf{Q})\nonumber \\
&&+\sum_{m'_2m'_3}\int d^3q''dp'dp''\nonumber
\\ &&\frac{U_{m'_2m'_3m_1}(\pi_2,\textbf{q}'',\textbf{Q})}{E-\frac{1}{m}(p''^2+\frac{3}{4}q''^2)}\frac{\hat{t}^a_{m_2m_3m'_3m'_3}(\textbf{p},\pi_1,z)}{E+i\epsilon-E_d-\frac{3q^2}{4m}}\nonumber
\\
\end{eqnarray}
After some simplification the integral equation transforms to this
equation:
\begin{eqnarray}
\label{Schmidth eq:1}
&&U_{m_1m_2m_3}(\textbf{p},\textbf{q},\textbf{Q})=U'_{m_1m_2m_3}(\textbf{p},\textbf{q},\textbf{Q})\nonumber \\
&&+\frac{2}{q}\sum_{m'_2m'_3}\int\limits_{0}^{\infty}dp'p'\frac{1}{E+i\epsilon-\frac{1}{m}(p'^2+\frac{3}{4}q^2)}\nonumber
\\ &&\int\limits_{|q/2-p'|}^{q/2+p'}dq''q''\bar{G}(q,q'',p')\int d\hat{q}''\delta(x''-x_0)\nonumber \\&&U_{m'_2m'_3m_1}(p''\hat{\pi}_2,\textbf{q}'',\textbf{Q})\hat{t}^a_{m_2m_3m'_3m'_3}(\textbf{p},p'\hat{\pi}_1,z)\nonumber
\\
&&-\frac{2}{q}\sum_{m'_2m'_3}\int\limits_{0}^{\infty}dq''q''\frac{1}{E+i\epsilon-E_d-\frac{3q^2}{4m}}\nonumber
\\ &&\int\limits_{|q/2-q''|}^{q/2+q''}dp'p'\bar{G}(q,q'',p')\int d\hat{q}''\delta(x''-x_0)\nonumber \\&&U_{m'_2m'_3m_1}(p''\hat{\pi}_2,\textbf{q}'',\textbf{Q})\hat{t}^a_{m_2m_3m'_3m'_3}(\textbf{p},p'\hat{\pi}_1,z)
\end{eqnarray}
In the above equation $\bar{G}$ which is always positive is defined
as:
\begin{eqnarray}
\label{Schmidth eq:1}
 \bar{G}(q,q'',p')=\frac{1}{-E_d-\frac{3q''^2}{4m}+\frac{1}{m}(p'^2+\frac{3}{4}q^2)}
\end{eqnarray}

$x''=\cos\theta''$ indicates the angle between $\textbf{q}$ and
$\textbf{q}''$ and $x_0$ is introduced as follow:
\begin{eqnarray}
\label{Schmidth eq:1}
 x_0=\frac{1}{qq''}(p'^2-\frac{1}{4}q^2-q''^2)=\frac{1}{qq''}(p''^2-\frac{1}{4}q''^2-q^2)\nonumber
 \\
\end{eqnarray}

\section{Summary and outlook}
\label{summary}

 In this paper we have formulated the Faddeev integral equations for
 calculating the photodisintegration observable of triton in a three
 dimensional approach. To this aim we introduced our basic states
 which contains jacobi momenta in vector forms as well as individual
 spin and isospin of each nucleon. So we have avoided to decompose
 angle states in terms of angular momentum states (partial wave
 approach)
 which is traditionally used to solve these kind of equations. The
 final integral equations are less complicated than the
 PW ones and are unique in number of the equations in all energies. We
 have also explained  about overcoming of the moving
 singularity in our work.

 The calculation of this observable using the AV18 potential is
 underway
and the results will be published soon.

Adding two and three body currents as well as three body forces  in
our calculations are other future major works. The same calculation
for radiative capture is also under consideration.

\section*{Acknowledgments}
This work was supported by center of excellence on structure of
matter, Department of Physics, University of Tehran.

\end{document}